# idwMapper: An interactive and data-driven web mapping framework for visualizing and sensing high-dimensional geospatial (big) data

Sarigai Sarigai [1,2], Liping Yang [1,2,3] *, Katie Slack [1,2], K. Maria D. Lane [1], Michaela Buenemann [4], Qiusheng Wu [5], Gordon Woodhull [6], Joshua Driscol[1,2,7]

* Corresponding author: lipingyang@unm.edu

[1] Department of Geography and Environmental Studies, University of New Mexico, Albuquerque, NM 87131, USA
[2] Center for the Advancement of Spatial Informatics Research and Education (ASPIRE), University of New Mexico, Albuquerque, NM 87131, USA
[3] Department of Computer Science, University of New Mexico, Albuquerque, NM 87106, USA
[4] Department of Geography and Environmental Studies, New Mexico State University, Las Cruces, NM 88003, USA
[5] Department of Geography & Sustainability, University of Tennessee, Knoxville, TN 37996, USA
[6] Posit PBC, Boston MA 02210, USA
[7] Woodwell Climate Research Center, Falmouth, MA 02540, USA

## Abstract

We are surrounded by overwhelming big data, which brings substantial advances but meanwhile poses many challenges. Geospatial big data comprises a big portion of big data, and is essential and powerful for decision-making if being utilized strategically. Volumes in size and high dimensions are two of the major challenges that prevent strategic decision-making from (geospatial) big data. Interactive map-based and geovisualization enabled web applications are intuitive and useful to construct knowledge and reveal insights from high-dimensional (geospatial) big data for actionable decision-making. We propose an interactive and data-driven web mapping framework, named *idwMapper*, for visualizing and sensing high dimensional geospatial (big) data in an interactive and scalable manner. To demonstrate the wide applicability and usefulness of our framework, we have applied our *idwMapper* framework to three real-world case studies and implemented three corresponding web map applications: iLit4GEE-AI, iWURanking, and iTRELISmap. We expect and hope the three web maps demonstrated in different domains, from literature big data analysis through world university ranking to scholar mapping, will provide a good start and inspire researchers and practitioners in various domains to apply our idwMapper to solve (or at least aid them in solving) their impactful problems.

## Abbreviations

The following abbreviations (ordered alphabetically) are used in this article:

ARWU
    Academic Ranking of World Universities

AI
    Artificial Intelligence
CSS
    Cascading Style Sheets
CSV
    Comma-separated Values
CV
    Computer Vision
D3
    Data-Driven Documents
DL
    Deep Learning
EDA
    Exploratory Data Analysis
GEE
    Google Earth Engine
GeoAI
    Geospatial Artificial Intelligence
GIS
    Geographic Information Science
HTML
    HyperText Markup Language
HTTP
    Hypertext Transfer Protocol
idwMapper
    **i**nteractive and **d**ata-driven **w**eb **M**apping framework
IREG
    International Ranking Expert Group
JS
    JavaScript
JSON
    JavaScript Object Notation
GeoJSON
    Geographical JavaScript Object Notation
ML
    Machine Learning
QS
    Quacquarelli Symonds
RS
    Remote Sensing
THE
    Times Higher Education
TRELIS
    Training and Retaining Leaders in STEM
UCD
    User-centered Design

WWW
World Wide Web

## 1. INTRODUCTION AND MOTIVATION

Cartography long ago made a transition from "maps" to "mapping" (Kitchin & Dodge, 2007). Interactive frameworks now dominate most mapping endeavors, and cartography is no longer available only to experts (Crampton, 2009). After a long history of exerting deterministic control over numerous aspects of daily life, cartography's increasing accessibility raises some interesting prospects for democratization (Akerman, 2009; Kim, 2015; Monmonier, 2010). Yet much is left hidden behind an easily accessible interactive mapping platform, including many specific constraints and decisions affecting data characteristics, accessibility, and visualization (Graham et al., 2015; Leszczynski, 2014). As the role of cartographer/mapmaker becomes ever more widely available and interactive, it is critical to consider the portals through which (geospatial) big data can be accessed and spatially rendered (Dalton et al., 2016).

In this paper, we consider data visualization, which has emerged as a primary tool for decision-making (Gotlib et al., 2021; Hallisey, 2005). As cartographic data platforms and data sets have proliferated, the frameworks for interaction with geospatial data have also multiplied. Recent work focuses on how to keep these frameworks open and accessible (Correll, 2019; Kelly, 2021; Zhunis et al., 2021). Visualization tools have the challenge of making legible meaning from data, regardless of who is using the tool. As Kitchin notes: "Making sense of data is always framed – data are examined through a particular lens that influences how they are interpreted." (Kitchin, 2014). In order for multiple users to develop actionable understanding and meaning from cartographic visualization, universal tools are needed that support visualization of data with high dimensionality. Key considerations in the design of visualization tools are speed, interactiveness, and user-centered characteristics.

Web map applications are powerful tools for better accessibility and approachability of data as they leverage recent advancements in interactive web browsers to support browser-based user interfaces and technologies without the need for computing environment setup before using the tools (e.g., downloading and installing required software) (Saia et al., 2022). Interactive web apps can also help researchers, practitioners, and policymakers generate insights from (geospatial big) data. They can help stakeholders and developers work more closely together to notice and solve problems. For example, interactive web app tools allow designers and users to spot some unveiling data entry problems during the design and development stages; without these tools, it may never be possible to notice the errors/typos during data entry. Designing and developing professional interactive web apps takes time, and very large volume and high-dimension big data characters embedded in (geospatial) big data significantly increase the challenges. *Systematic frameworks that can guide users to develop professional web mapping applications remain limited; systematic frameworks here refer to those considered professional design principles such as information visualization and cartographic design principles plus technique specifics such as programming.* We work towards filling this gap by proposing, designing, and implementing a professional interactive and data-driven web mapping framework named *idwMapper*, with three real-world motivated web map applications, to widely demonstrate the universal needs and usefulness of such interactive and data-driven web mapping framework.

Here we provide a roadmap to the rest of the paper. Section 2 outlines some background to prepare our readers to understand the paper. Sections 3, 4, and 5 are the core of the paper. Section 3 introduces the *idwMapper* framework, its overall architecture, its core coordinated-view visualization (Section 3.2), and visual components and their associated data types (Sections 3.3 and 3.4). Section 4 introduces design principles suggested in our *idwMapper,* including information visualization and cartographical design principles, detailed in Sections 4.1 and 4.2. Section 5 presents three case studies with different domain focuses to demonstrate the wide applicability of our web mapping framework *idwMapper* (detailed in Sections 5.2, 5.3, and 5.4, respectively). Note that as the three case studies designed and implemented in this paper are stand-alone use cases of our proposed idwMapper, picking the usage case(s) based on our audience's interest will not affect the readability. The paper concludes in Section 6 discussing key challenges and opportunities. As there are some acronyms in this paper, we provide a full list of abbreviations right after the abstract and before the introduction for readability.

## 2. BACKGROUND

In this section, we review some previous work that bears on our contribution, geospatial big data, information visualization, geovisualization, and web frameworks.

### 2.1 Geospatial big data

Big data is substantial in volume, high in velocity, diverse in variety, and exhaustive in scope (Kitchin, 2014). Geospatial big data is a significant and rapidly increasing part of big data, with a yearly growth rate of at least 20% (Lee & Kang, 2015). Beyond the shared attributes of big data stated above, geospatial big data is often highly dimensional, and the interaction among variables is complex; thus from which human vision cannot easily get insights (MacEachren & Kraak, 2001). Large volumes and a wide range of high-quality geospatial data are more obtainable for researchers (Dykes et al., 2005). However, leveraging geospatial big data to implement various environmental and societal applications is challenging (Parente et al., 2019). Computational and visual approaches can be integrated to improve knowledge construction from geospatial data, as the integration can uncover the hidden patterns and relationships in such complex data (MacEachren & Kraak, 2001). Visual platforms also enable intuitive analysis of spatio-temporal data in a multidimensional manner and provide access to cartographic displays as well as tabular and graphic displays, which enable varying levels of aggregation (Lee & Kang, 2015).

### 2.2 Information visualization and geovisualization

Information visualization is the process of representing data visually (Gershon et al., 1998). Dashboards along with different charts are common examples of information visualization. Information visualization allows researchers to explore, analyze, interact, and visualize data to draw insights from abstract data by depicting an overview and making visualizing hidden connections possible. Geovisualization is a field that draws upon methods and approaches from many disciplines, including geography, cartography, information visualization, image analysis, exploratory data analysis (EDA), and geographic information science (GIS). Maps provide insight generation from geospatial patterns; cartography is the art and science of designing, producing, and using maps (Liu et al., 2014). Technological development transformed how maps were made, used, and shared (Cartwright, 1999). With the development of technology and the introduction of the World Wide Web (WWW), the way of acquiring, managing, analyzing, and cartographically representing geospatial data has substantially changed (M. J. Kraak, 2001). Web maps are interactive displays of geographic information in the form of a web page that allow users to tell stories, construct knowledge, and answer questions (Dorman, 2020; MacEachren, 1995). Web cartography and web mapping have dominated the way of producing maps, replacing paper maps to some degree (MacEachren & Kraak, 2001). The most significant advantage of web mapping is its accessibility; it can be easily reached by anyone who has an internet connection through an internet browser (R. E. Roth et al., 2015). Over the past decade, the integration of web maps into websites has become essential; this can be attributed, among other factors, to significant advancements in geographic data delivery methods and the accumulation of extensive collections of such data by both private and public entities (Krol, 2019).

   With the increasing amount of multi-dimensional geospatial information, the difficulties of sense-making and communication of number-rich datasets are also rising (R. E. Roth, 2015). Traditional representation methods that only use static maps or simple graphics to visualize and explore data are no longer effective. New challenges and opportunities have been posed to geovisualization with the development of interactive computer tools, interface design, and related technologies (Liu et al., 2014). Web maps can increase the accessibility of geospatial big data, as web maps can be used to read the data from and interact with, but they also contain other information visuals such as text, tables, graphs, and charts. The use of maps and graphics for geovisualization and information visualization has been accepted as a significant way of data storytelling. Maps can give spatial structure to oral, written, and audio-visual forms of storytelling, and often are combined with graphics, images, videos, and text to provide a deep account of people, places, and events (R. E. Roth, 2021). Combining web maps and visuals makes the information more accessible, actionable, and meaningful. Well-designed maps, especially web maps,  have become an interface combined with other visuals that can be leveraged for easy information access and a *productive knowledge construction process* (MacEachren, 1995; MacEachren & Kraak, 2001). Web mapping applications allow users to interconnect and distribute geospatial resources often stored on the WWW as its "database", with support of their two major characteristics – interaction and dynamic (Dykes et al., 2005). Interaction as a major function of modern maps changed maps in both quantitative and qualitative ways. It not only changed the way how maps are distributed and the number of users that can access them, but also, more importantly, changed the way users understand, discover, and explore geospatial data using interactive maps, and consequently make decisions (R. E. Roth, 2013). Web mapping applications became an essential tool to visualize, analyze, explore, and understand geospatial data.

## 2.3 Web frameworks and challenges

The widespread use of digital interactive mapping and analysis technologies allows a significant shift in the use of maps in various facets of society and extends beyond the realm of professional cartographers (Nelson et al., 2022). Crowdsourcing and citizen science in cartography have also fastened the "rise of amateur" -- a shift from map readers to map makers (Dodge & Kitchin, 2013); this also refers to user-cartography (A. J. Kent, 2018). Most of the recent information visualization studies focus on empirical methodologies and applications (Liu et al., 2014). Many web map applications and dashboards have been created. (Big) companies released their own visualization platforms and dashboards, such as ArcGIS Dashboards[1], Shiny[2], Voila[3], Dash[4], Streamlit[5], Panel[6], and Bokeh[7]. Those dashboards are often either subject-specific or maps are an add-on like any other chart, and most importantly, they often lack cartographic design considerations. (Lan et al., 2021) also indicated that although the dashboards and maps are helpful for both experts and the general public to understand different research problems such as climate change issues easily, researchers questioned the wide usage of dashboards and the design and efficiency of web maps. Web maps have advanced map storage and sharing, consequently giving rise to user-cartography. However, maps used in dashboards can be made by individuals with limited cartographic knowledge, which may lead to poorly designed and inefficient maps (Lan et al., 2021; MacEachren, 1994, 1995). Moreover, researchers emphasized that cartographic techniques such as scales and symbols can be a big factor in distorting the information delivery to end users if used improperly (Dodge & Kitchin, 2013). It can thus lead to the danger of user-cartographers disseminating misleading information via maps due to their lack of domain knowledge. While interactive maps look great, ensuring they "work" successfully for the target users remains a challenge for designers and developers (MacEachren, 1995). Researchers emphasized the importance of developing the students' and users' understanding of (spatial) data, as spatial data are just one possible representation of reality (Bearman et al., 2016). There is a risk of developing recipe-type approaches to the practicals, where users just apply the data without understanding what they mean (Elwood & Wilson, 2017). Further, although web mapping is conceptually fairly straightforward, the techniques for actually implementing maps on the web can be detailed and very technical, and computer programming skills are often required to develop web mapping applications. Thus, open educational resources that combine cartographic design guidance with web mapping technologies are needed to train a new generation of cartographers, students, researchers, and the public. To address the challenges stated above, *there is a strong need to develop useful frameworks with implied/suggested information visualization and cartography design principles, along with geography knowledge.*

## 3. THE *idwMapper* FRAMEWORK

We designed and developed *idwMapper*, an interactive and data-driven web mapping framework for web map application development towards sensing high-dimensional geospatial big data. *idwMapper* provides a combination of theoretical and practical guidelines, plus we have specifically implemented three case studies web map tools to demonstrate the applicability of the framework and to inspire our users to create their own professional web mapping applications with minimal design and coding efforts. More specifically, *idwMapper* leverages coordinated-view visualization (Section 3.2) and cross-filtering to handle high dimensional geospatial data with complex variables, and importantly *idwMapper* implies professional design guides from information visualization (Section 4.1) and cartographic design (Section 4.2) perspectives.

### 3.1 *idwMapper* framework overall architecture

Figure 1 provides the overall *idwMapper* workflow, which includes two major components, and they are iteratively linked. One component is to obtain and preprocess data, and the other is to design and implement a web map application using the *idwMapper* framework. Note that there is no specific order in which components go first, as that all depends on different scenarios. For some cases it might start with a small set of data and the other cases might start from analyzing user needs and then start to collect data. The data can be manually collected and/or web-scraped. The *idwMapper* recommended data formats are comma-separated values (CSV), JavaScript object notation (JSON), or geographical JavaScript object notation (GeoJSON). When designing a web application, its target users and their needs are the most important things to consider. User needs are then translated to the web application utility requirements. Specifically, based on user needs, a set of map functionalities and visual components recommended by our *idwMapper* framework (detailed in Section 3.3) can be selected. After having the final formation of needed functionalities and visuals, the HyperText Markup Language (HTML), JavaScript (JS), and Cascading Style Sheets (CSS) codes can be adapted from the provided *idwMapper case study* coding examples *(detailed in Section 5)*, to build a web application.

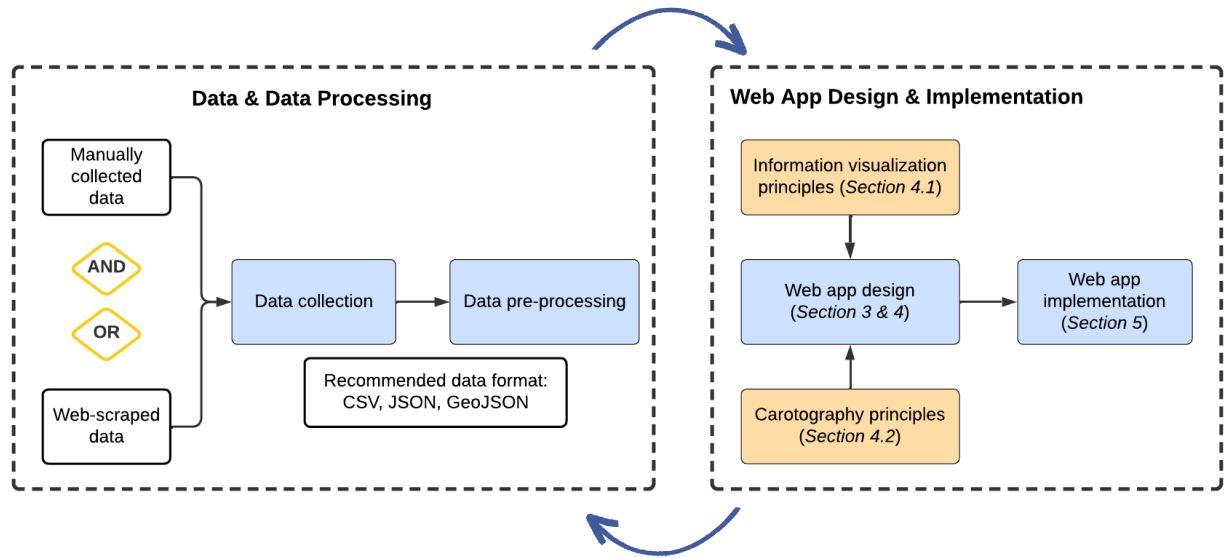

Figure 1. The overall workflow of *idwMapper*. The data source can be manually collected and/or automated via web scraping, based on different web map application needs. For example, in Section 5, our illustration case study map iLit4GEE-AI (Section 5.2) and iTRELISmap (Section 5.3) data are manually connected by some authors of this *idwMapper* paper, and iWURanking (Section 5.4) data are obtained via web scraping. To help and/or inspire people in different domains to develop professional web maps with minimal coding, we introduced information visualization and cartography principles (detailed in Sections 4.1 and 4.2 below).

## 3.2 Coordinated-view visualization

**Coordinated-view visualization is the soul of our *idwMapper***, as it is the cornerstone that makes *idwMapper* possible for sensing high-dimensional (geospatial) big data. Interactivity importantly differentiates web apps from traditional maps, and coordinated-view visualization enabled systems essentially improve interactivity (Robinson, 2017). Coordinated-view visualization, which is an exploratory visualization, gives users the flexibility to achieve their desired goals using different paths as multiple views are simultaneously linked together. Multiple views in information visualization refer to different dimensions of the same dataset represented in multiple visuals (e.g., charts and tables), and coordinated-view visualization allows those multiple visuals to be linked together simultaneously (i.e., making selection changes in one visual update the other visuals). Coordinated-view visualization allows geospatial scientists to link different datasets together in any combination at different scales and dimensions to reveal hidden patterns and relationships from geospatial big data (Dykes et al., 2005; M.-J. Kraak & Ormeling, 2020; Scherr, 2008). Users can also gain a very deep understanding of spatial data through the selection and cross-filter enabled by coordinated-view visualization. Coordinated-view visualization is essential, especially when data are huge in volume and complex in structure. With traditional geovisualization views, such as adding a map and other charts and graphs to an interface that has no view connection with each other, it is challenging to extract hidden patterns from high-dimensional complex datasets.

### 3.3 Visual components and how and when to use guide

*The idwMapper* comprises multiple coordinated-views through cross-filtering (detailed in Section 3.2), which include web maps, charts, and a data table (as illustrated in Figure 2). This section introduces the *idwMapper* layout and the commonly used charts along with suggestions for when and how to use them appropriately.

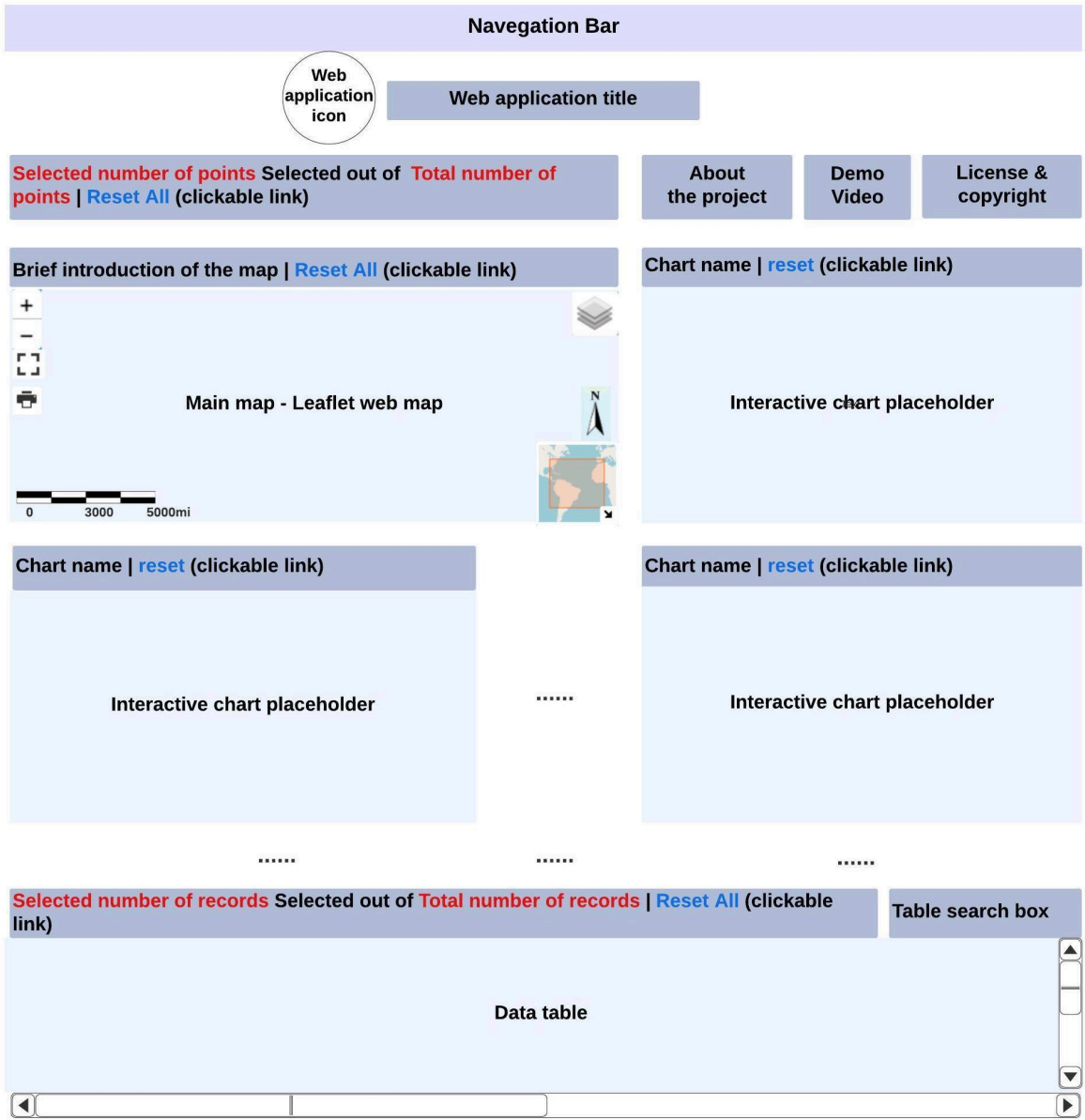

Figure 2. The *idwMapper* framework layout, which contains four major components: *(1)* main map that includes essential elements of a map and additional elements such as mini-map and print function; *(2) charts.* The chart placeholders in the example layout only serve as examples; the number and size of the charts will vary from case to case; *(3)* data table; *(4)* elements such as navigation bar, web app icon and titles as well as instructions/videos. The default data table when the map is initially loaded displays the whole dataset – it will be updated when other charts and/or map filters.

### 3.3.1 *Main map component*

The web map is essential for *idwMapper* as it serves as the central map visualization interface for high-dimension of geospatial big data. The main web map in *idwMapper* is implemented using open-source Leaflet.js. In *idwMapper*, the leaflet marker map is used to create the web map and geographically visualize the data. The marker cluster plugin of Leaflet provides professional animated marker clustering functionality that allows users to effectively visualize the data having the same/close range of latitude and longitude. Pop-ups deliver detailed information on each point in an efficient way without a heavy cognitive load (see an example of pop-up in Figure 3 (e)). Different basemap options allow users to explore geographic data from different perspectives (e.g., satellite or street map). Some auxiliary functions such as print, and view in full screen provide flexibility to users to save/export the whole data or a subset of data they are interested in.

### 3.3.2 Charts

In a geovisualization context, other cartographic products such as charts have received demanding attention because they build knowledge interactively with existing expertise and knowledge base of maps (Dykes et al., 2005). Interactive charts interact with the visual system easily and pass information fast. Researchers also noted that charts are helpful to allow users to easily think over the information they want to take away without being too overwhelmed (Knaflic, 2015). There are two categories of charts: static and dynamic (M. Chen et al., 2009). The charts used in the *idwMapper* are **dynamic**. Different charts have different ways of conveying large volumes of data in an easily understandable format to address users' visualization needs. *idwMapper* includes the following charts: bar charts, pie charts (or donut charts), line charts, row charts (i.e., horizontal bar charts), row charts with x-axis scroll, and sunburst charts. Charts need to be carefully selected to properly convey information for certain types of data. Below we provide some guides for when to use which chart.

   **Pie/donut charts** are commonly used to show parts of a whole relation of categorical variables (Knaflic, 2015). Donut charts are useful because they are more space-efficient than pie charts. For example, the legend can be placed inside the inner circle space of the donut chart or just left blank for white space, which helps reduce users' information overload. Traditional static donut charts may not be visually effective especially when there are many categories. Donut charts designed and implemented in our *idwMapper* have clickable legends to avoid the visual effectiveness problem. When there are many categories, the scrollable legend allows users to quickly navigate to and select based-on their interests.

   **Bar charts and row charts (horizontal bar charts)** are used to present categorical, discrete, and continuous variables. In *idwMapper*, we use both vertical and horizontal bar charts. Horizontal bar charts are suggested when category names are long and there are more categories because they can be added with x-axis and y-axis scroll functions more feasibly. The brushing function is enabled in some charts to make the data visualization more effective. Brushing is the interactive function that defines a selected region, either one or two dimensional, through a pointing gesture such as clicking and dragging the mouse (Crampton, 2002; MacEachren, 1995; Nöllenburg, 2007). The advantage of the brushing function in a bar chart is that multiple continuous bars can be efficiently selected through a simple single mouse drag; whereas a bar chart without the brushing function requires multiple clicks for multiple selections. When the data span is large it is cumbersome for users to select multiple bars, so a brushing function is needed. However, once the brushing function is enabled, the individual bar selection will be disabled. Thus, when the data span is not large, the brushing function is not very necessary.

**Sunburst charts** are useful when dealing with multi-levels data (i.e., data hierarchy). They complement other proportional charts that are not able to deal with hierarchy such as pie charts. Sunburst charts can visualize hierarchical data structures such as those that have order level and organizational structure. More specifically, a sunburst chart includes an inner circle surrounded by rings of deeper hierarchy levels. The angle of each segment is either proportional to a value or divided equally under its parent node. Our *idwMapper* is designed to visualize high-dimensional geospatial data, and so the sunburst chart is helpful to visualize some complex and layered data. Sunburst charts have the advantage of clearly showing the dimension of depth to each parent branch. However, similar to pie charts and donut charts, when there are too many categories in each branch, it might not be able to clearly present every single category. To solve this problem, in *idwMapper*, we also added scrollable legends functionality to all sunburst charts. Users can easily select and locate the slice that they are interested in looking at by scrolling through the interactive legends.

**The zoom and focus line chart** is a combination of two line charts. Features like zoom and focus allow users to focus on sections of the line chart. More specifically it lets the users zoom in and focus on a selected range of data in the charts. For example, if a user wants to focus on universities that have an overall score from 90 to 100, they could move the brush to select the range of 90 to 100 in the zoom chart (the small range chart below the main chart), the focus chart will then zoom in to these selected scores (see Figure 10 in Section 5.4.2 for an example). It can be useful when a dataset range is large, so the main line chart provides the overview, and the zoom and focus feature provides detailed information. Zoom and focus line charts provide users an interactive opportunity to see more details within a selected data window.

### 3.3.3 Word clouds

Word clouds are powerful to visualize big textual data because they can help people intuitively grasp the major topics from big data without needing to read the entire large data set (Kalmukov, 2021). For example, word clouds were implemented in (Bertolino et al., 2018) to identify the topics of highest interest to the authors from the titles of the papers published at various conferences. Program committee chairs are suggested to use word clouds generated from abstracts of all submitted papers to avoid unbalanced subjects (Kalmukov, 2021). In *idwMapper*, a word cloud is a tool to enable dynamic information delivery by coordinating with the main map, graphs and charts, and the data table; which means the default word cloud reflects the whole data set and it will be updated to reflect the sub dataset cross-filtered by the map and/or other charts and graphs. The importance of tags' visual properties was emphasized in (Bateman et al., 2008), including font size, weight, color, intensity, and position. In *idwMapper*, the visual properties are considered to assist the audience in better visualization and information gain.

### 3.3.4 Data table

Data table is one important overview component in our *idwMapper*. Tables interact with the user's verbal system; it, sometimes, is the easiest way to convey information to our audience (Knaflic, 2015). Tables are also the most inclusive tool to present data because many columns can be easily added to them to enrich the semantic context of a web app, while other charts only allow limited information. The data table designed and implemented in *idwMapper* has several useful functions that make the information visualization effective. For example, the table in *idwMapper* is linked with all other charts and maps -- specifically the data table is updated dynamically by multiple other maps and/or charts through crossfilters. The search function also helps users find the information they are interested in quickly by using their entered keyword in the search box. The sorting function also allows the user to pick up information easily.

### 3.4 Data types for visual components

It is essential to know when to use the proper charts for the corresponding data type. Tables 1 and 2 below summarize such types of guidance. NA in both tables refers to not applicable.

Table 1. The data type and functions of map and map-related components. NA refers to not applicable.

| Map/map related component | Data Type | Legend | Events |
|---|---|---|---|
| Marker map | Location data (point) | Marker map legend | NA |
| Basemap | Tile layer | NA | Turn on/off |
| Pop-up | Text, image, audio data | NA | Scrollable, content can be clickable |

Table 2. The data type and functions of other components. Brushing is the interactive function that defines a selected region, either one or two dimensional, through a pointing gesture, such as clicking and dragging the mouse (detailed in Section 3.3.2). Note that all charts/graphs will be updated according to those (combined) filtering from those charts that have filterable functions, such as the map and the row charts. Here "array filter/dimension tag" means a column can have values that are separated by commas. For example, one row value for a column could be "A,B, C" and the other row value be "A,B,D"; if we use the column to generate a chart, we are able to use single value "A", "B", "C", "D" and or any combination of which to filter the data set.

| Chart/graph | Data Type | Legend | Events |
|---|---|---|---|
| Bar chart | Categorical data | NA | Brushing, clickable, array filter/dimension tag, filterable |
| Line chart | Continuous data | NA | Brushing, clickable, filterable |
| Pie chart/Donut chart | Categorical data | Clickable, scrollable | Clickable, array filter/dimension tag, filterable |
| Row chart (i.e., Horizontal bar chart) | Categorical data | NA | x-axis scroll, y-axis scroll, brushing, array filter/dimension tag, filterable |
| Row chart with x-axis scroll | Categorical data | NA | Clickable, x-axis scroll,  y-axis |
| (i.e., Horizontal bar chart with x-axis scroll ) | | | scroll, array filter/dimension tag, filterable |
| Sunburst chart | Categorical data, hierarchical data | Clickable, scrollable | Clickable, filterable |
| Word cloud | Textual data | NA | Clickable |
| Data table | CSV data | NA | Searchable, paginated, scrollable (horizontally), clickable, ordering |

## 4 DESIGN PRINCIPLES IMPLIED IN idwMapper FRAMEWORK

Visual selection can sometimes be challenging, but it is also the key step of the information visualization process. Text, graphs, and data tables are three visuals we suggested as essential visuals besides a map to add to a web map application. The detailed introduction and applied data type to each graph are listed in Sections 3.3 and 3.4. Users are recommended to choose an appropriate visual type that best fits their data and can accomplish the visualization goal in order to communicate with the data most effectively. In the following two sections, we provide information visualization and cartographic design principles to make their web map visual components professional and function as expected.

### 4.1 Information visualization principles

The primary goal of information visualization is to enable users to gain insightful understanding (C. Chen, 2010). The key to success is to allow users to gain useful insights from the visuals with ease of use. Below we provide information visualization design principles implied and emphasized by our *idwMapper* framework.

#### 4.1.1 *idwMapper* suggests avoiding clutters and removing any unnecessary elements

Clutter is one big enemy of information visualization (Knaflic, 2015). Clutter adds a cognitive overload to our audience and makes the information delivery inefficient. Fighting clutter is like fighting weeds (Zinsser, 2001). Our *idwMapper* suggests carefully examining every element being put onto the web map layout. Users will find a surprising number that do not serve any good purpose.

#### 4.1.2 *idwMapper* emphasizes a user-centered design

Knowing the audience is crucial to a good information visualization design. **User-centered design (**UCD) is a method that emphasizes an early and active focus on user's needs and iterative refinement when conceptualizing and implementing a user-friendly and useful interface (Koh et al., 2011). UCD emphasizes high usability and utility of an interface (R. E. Roth & MacEachren, 2016). Usability describes the ease of using an interface to complete the user's desired tasks. Utility describes how useful an interface is to allow the users to complete their objectives (Janicki et al., 2016). The three U's of interface success proposed in (R. Roth et al., 2015) form a triangular loop relationship that each of the three components is contingent upon refinements to the prior; they emphasize the importance of considering all three Us (user, utility, and usability) within the three U's loop. To create successful and user-friendly UCD visualization applications, *idwMapaper* recommends UCD and considers all three essential elements.

### 4.1.3 *idwMapper* suggests choosing an appropriate chart type

It is important to choose an appropriate chart type for representing a specific type of data. It helps the audience discover actionable insights. The chart selection should concern the story users are trying to tell, the purpose of the visualization, the problem supposed to be solved by the visualization, and the data you have. A brief introduction of each chart suggested in *idwMapper* is described in Section 3.3 and the data type supported by different charts is listed in Table 2.

### 4.1.4 *idwMapper* suggests keeping the important things highlighted and others simple

By facilitating the comprehension of the intended information by emphasizing the most significant and fundamental information, users are enabled to grasp the entire story readily. For example, providing a descriptive and informative title to a web application enables users to immediately apprehend the central purpose of the application and retain this comprehension while browsing. By contrast, some information needs to be simplified to allow users to visually derive conclusions from data easily. Visualizations that are ineffective demand users' greater degree of cognitive effort to process information.

### 4.1.5 *idwMapper* emphasizes using color wisely

Color selection is an essential process for information visualization and map making. An effective color selection improves the communication of geographic data (Brewer, 1994). In explanatory analysis, the use of color and text is one way to focus on the story instead of just showing data. In information visualization, color is one of the most important tools for attracting users' attention (Knaflic, 2015). A common mistake of ineffective visualization is using color for the sake of beauty. Although the aesthetic of visuals is important, leveraging color selectively as a strategic tool to highlight the essential parts of visuals is even more important. In *idwMapper*, we suggest our users utilize color wisely, as color always conveys something no matter whether we know it explicitly or not. *(1) idwMapper values the consideration of color blindness.* Research shows that roughly 8% of men and 0.5% of women are color blind. The most common difficulty they have is distinguishing between shades of red and green. Thus, we suggest users do not use shades of red and green together. Moreover, there are certain color palettes that are simulated by authors via software to optimize for color-blind individuals (Wong, 2011). Diagnostic tools such as ColorBrewer [8] are designed to evaluate the robustness of individual color schemes and provide options to select color-blind safe schemes. Simulator tools such as Color Blind Vision Simulator [9] allow users to simulate their image by using the different types of color blindness to check whether their image is color blind safe (we used those tools to check color blind safe in our use cases in Section 5). *(2) idwMapper suggests using a consistent color scheme as this helps audiences understand the information, making it easier to interpret later graphs and reducing mental fatigue.* When keeping the color layout consistent, it is easier for audiences to notice the real changes that we want to show by leveraging color. So use colors for meaningful reasons (Knaflic, 2015). *(3) idwMapper emphasizes careful selection of color by considering the meaning behind color.* For example, in our use case study map, iTRELISmap (see Section 5.3), as it is a female scholars group, we chose purple as the main color hue of the web map.

## 4.2 Cartography principles

In this section, we provide the cartography principles implied and emphasized by *idwMapper*.

### 4.2.1 *idwMapper* emphasizes the importance of map aesthetics

Art and science are integral parts of cartography (Krygier, 1995). Maps often aim to provide geospatial insights to their users. The main goal of mapping is to accurately represent space with emphasis on accuracy, clarity, and usefulness for both professionals (e.g., cartographers) and amateurs. Aesthetic aspects (e.g., design, color, and typography) also play a crucial role in attracting and retaining viewers' attention. No matter the kind of purpose the map users and mapmakers have for creating certain maps, maps need to be both functional and aesthetic to help users explore data and extract information. The common criteria for map aesthetics are *harmony, composition, and clarity* (A. Kent, 2013). Harmony indicates the interactions between different map elements; see if the map elements could work together to provide better visual effects and help users communicate better with the information. The composition is similar to the page layout; it deals with the spacing, weighting, and placement of elements included. Clarity refers to information delivery efficiency. Maps must be clear enough to let the user easily gain and recognize information from the map. Different types of map functions are helpful for users to grasp information from the map. Map functions are discussed in Section 4.2.2 directly below considering essential elements. The basic functions of web maps are the same as traditional maps. Due to the nature of web maps (e.g., interactiveness, overlaid multiple layers and real-time updates), additional functions and design principles are needed to make web maps functional and useful. Beyond the basic criteria for web maps, the functionality is closely related to its interactiveness, accessibility, speed, and user-centered considerations.

### 4.2.2 *idwMapper* recommends designing and making functional maps

*idwMapper* is a framework that aims to sense high-dimensional geospatial data, so maps are a major component. Maps are useful for exploring patterns, comparing and analyzing different environmental and social relations of place, and consequently helpful for decision-making (MacEachren & Kraak, 2001). See below for five suggestions for how to design and create a functional map. (1) *A functional map includes all principal map elements: map title, legend, scale bar, and north arrow*. The map title has to be included to provide the users with the map's overview and define the map's context. A legend is necessary to let the user know information about the representational elements in the map. If there is no legend, even if the map is very comprehensive, users might not be able to know which points/lines/polygons represent what data. Scale is one of the most important factors in geography and thus of a map; it is an abstract symbolic representation of place, as reality can never be displayed on a map. A scale bar has to be added to the map to indicate the relative measurement of a map compared to the real world. A north arrow is often essential to give the map a directional orientation of the real world (Hasan, 2020). (2) *A functional map has a carefully designed page layout that considers interface redundancy*. A web page needs to be balanced; information/interface redundancy needs to be considered together with the page layout as every single element in the map can add cognitive load to users' brains and consequently lead to bad information delivery. Redundancy here refers to strategic repetition of content in different formats (e.g., using graphics along with descriptive text). Redundancy of graphics and words is most beneficial to accommodate individual differences in cognitive abilities (Tindall-Ford et al., 1997). (3) *A functional map considers* *visual preattentive attributes to draw users' attention quickly and deliver information easily.* It includes the choice of the following: (1) color (detailed in Section 4.1.5 above), symbols, and icons, (2) line width, and (3) text size. These preattentive attributes are also the key elements for making an aesthetic map. (4) *A functional map should present the information it intends to deliver.* A good delivery of a map's purpose matters and could make the map functions right. In other words, knowing the purpose before making a map is helpful to create a functional map. (5) *A functional map uses an appropriate map projection.* An inappropriate map projection can confuse users and cause very misleading information (detailed in Section 4.2.4 below).

### 4.2.3 *idwMapper* suggests applying cartographic design principles

Four cartographic design principles were summarized and applied by (Janicki et al., 2016) to design and develop effective and aesthetically pleasing web mapping applications based on the principles from (Shneiderman, 2003) and (R. E. Roth & Harrower, 2008). Cartographic design principles for interactive maps include (1) overview first, detail on demand (Shneiderman, 2003); (2) map browsing flexibility (R. E. Roth & Harrower, 2008); (3) interface redundancy; and (4) interaction flexibility. Overview first, detail on demand is helpful to avoid information overloading the users. For example, the pop-up of a marker in the web app provides more information about the location to the users only when they click and select the location. Redundancy involves presenting content through multiple formats/ways, like using both visuals and descriptive text to convey the same information. This redundancy, combining graphics and words, is particularly advantageous for addressing variations in cognitive capacities among individuals (Tindall-Ford et al., 1997). In other words, some users might perceive visuals quicker than text, and vice versa. *idwMapper* delivers the same information through maps, graphs/charts, and data tables, which benefits users with different cognitive abilities. Interaction flexibility is the way to provide the user the ability to complete a task by taking different paths allowing a more flexible interface (Janicki et al., 2016). For example, in *idwMapper*, table search is a way to find information and the user's selection of chart combinations is another way of finding the desired information.

### 4.2.4 *idwMapper* emphasizes using a proper map projection

As stated in Section 4.2.2, a functional map uses a proper map projection. *idwMapper* emphasizes using a proper map projection according to the web map purpose. The three case studies applying *idwMapper* in Section 5 use Web Mercator projection as it is the most commonly used online map projection (note that the projection in the case studies can be changed to other projection according to the needs of other specific map purposes); and Web Mercator works well for our case studies, as the maps do not involve measurements such as areas and distance. Web Mercator is used by many online map providers (e.g., Google Maps, ESRI, OpenStreetMaps, Mapbox) (Battersby et al., 2014). The Mercator projection is a cylindrical and conformal projection that preserves local angles around points. The preservation of this property allows it to be suitable for certain purposes (e.g., navigation) but it distorts other properties (e.g., area). While the Web Mercator projection and the traditional Mercator projection have notable mathematical differences, at a global scale, it is not possible for people to identify the difference between the two projections. Web Mercator also simplifies the standard Mercator projection by mapping the Earth to a sphere, which allows for simpler (and therefore quicker) calculations; it supports web map service requirements to index "the world map" and allow for continuous panning and zooming to any area, at any location, and at any scale (Battersby et al., 2014).

## 5. CASE STUDIES APPLYING *idwMapper* FRAMEWORK

To demonstrate the wide applicability of *idwMapper* for developing interactive web apps for sensing geospatial big data, we designed and developed three case studies: iLit4GEE-AI, iWURanking, and iTRELISmap. This section will introduce the implementation overview, including main libraries used in these three case studies developed using *idwMapper* (Section 5.1) and the three case studies (Section 5.2 to Section 5.4).

### 5.1 Implementation overview

All three case studies' web apps were implemented in JS. The essential JS libraries we used are *D3 (Data-Driven Documents)*[10], *Crossfilter*[11], *and DC.js (Dimensional Charting JavaScript (JS) library)*[12], plus *Leaflet.js*[13], as well as some other JS libraries *DataTables*[14], *jQuery*[15], and *Bootstrap*[16]. (1) **D3** is an advanced and powerful data-driven JS library designed to create interactive, dynamic, and responsive data visualizations. (2) **Crossfilter** is a JS library for exploring large multivariate datasets. Crossfilter supports extremely fast (< 30ms) interaction with the coordinated-view, even with datasets containing a million+ records. The coordinated-view visualization and cross-filtering function that makes *idwMapper* powerful in conveying complex insight to audiences are implemented by Crossfilter. (3) **DC.js** is the core JS library used to implement the proposed *idwMapper* framework. It is an important and efficient JS library built upon D3 for interactive data analysis, especially useful for high-dimensional data. More specifically, DC.js allows complex data visualization to be fastly rendered and has a designed dashboard with various built-in charts and maps. DC.js is built to work with Crossfilter to handle coordinated-view visualization (Section 3.2) for high-dimensional data. (4) **Leaflet.js** is a widely used open-source JS library for creating interactive web maps due to it being lightweight, relatively simple, and very flexible. It works across all major desktop and mobile platforms. Also, Leaflet.js functionality can be easily extended through Leaflet plugins. All three case study web maps are hosted on GitHub and are open to the public (see DATA AND CODES AVAILABILITY STATEMENT right before the REFERENCES for more details).

### 5.2 iLit4GEE-AI web map app
#### 5.2.1 Motivation

A review paper summarizes the existing knowledge on a given topic in a research field and thus helps readers find the gaps in a particular topic to advance the field. Review papers often need to provide comprehensive and systematic summarization synthesis of existing papers; depending on the domain topic, often the number of papers to be reviewed is substantial. Researchers emphasize the importance of adding good figures and new technology to make the literature review process efficient and make the insight-finding process straightforward and eventually provide clear communication to readers (Tay, 2020). Most existing review papers that used visualization technology intensively do so by adding more advanced graphs or using the interaction application; these are often bibliometric reviews as they analyze extensive published research (Paul & Criado, 2020). There are few papers and tools that take advantage of advanced interactive web map applications and consider geovisualization. iLit4GEE-AI [17] is the one of the three web map application case studies designed, implemented, and developed by applying our *idwMapper* framework. It is the accompanying interactive web app for the review paper in (Yang et al., 2022). With this web application, we were able to make the substantial number of papers we reviewed (200 total) transparent. iLit4GEE-AI allows our readers to intuitively retrieve relevant Google Earth Engine (GEE) with Artificial Intelligence (AI) literature that fits their needs. For example, a user can quickly filter published articles that used Random Forest models and F1-score for wetland mapping. The accompanying web application played an important role in getting insights from the 200 review papers.

### 5.2.2 Data collection, preprocessing, and data structure

The data used in this web application were collected and extracted by the first three authors of the review paper (Yang et al., 2022). During the literature review phase, we initially identified over 500 papers relevant to GEE. Subsequently, we conducted a systematic search using the following criteria: (1) keyword search on Google Scholar: We utilized the keywords "Google Earth Engine" in conjunction with "machine learning," "deep learning," or "computer vision" for our literature search; (2) references tracking: We examined the references cited in recent GEE reviews, specifically the "References" sections of those papers. Additionally, we tracked newly published papers from the past two years that cited existing GEE review papers on their Google Scholar pages. It is important to note that our search was confined to research articles published in English and included in peer-reviewed journals or conference proceedings. After a rigorous selection process, we identified a total of 200 highly pertinent papers. Papers were excluded if they solely utilized GEE for remote sensing data download or if they did not incorporate AI, encompassing computer vision (CV), machine learning (ML), and deep learning (DL). Detailed data structure and data type explanation is available on this web page (https://geoair-lab.github.io/iLit4GEE-AI-WebApp/abbrlist_dataExplaination.html) on the iLit4GEE-AI web map app.

### 5.2.3 Web app design and implementation

In iLit4GEE-AI (https://geoair-lab.github.io/iLit4GEE-AI-WebApp/index.html), we interactively visualized and analyzed 200 reviewed papers. We collected the following data from each of the reviewed papers: paper title, method or application-oriented, application focus, paper's Google Scholar URL, Google Scholar cited by URL, cited by number as of data entry date, names of authors, name of the first author, first author's Google Scholar profile, first author's institution, country of first author's institution, the latitude and longitude of first author's institution, journal name, journal official site URL, published year, abstract, PDF link of paper open-access version, keywords, type of remote sensing data used, study area, method (macro), method (detailed), models compared, compute (cloud platform), compute (offline hardware), other software, and evaluation metrics. We carefully followed the design principles suggested in *idwMapper* (detailed in Section 4) for chart selection to visually present the data. Specifically, we use pie charts to visualize the country of the first author's institution, published in journal, application focus, study area, RS data type, method/application, method (macro), method (detailed), hierarchy of methods, evaluation metrics, models compared, computed on cloud (see Figure 3). We chose a bar chart to visualize the year of publication (Figure 3b) and row charts to visualize keywords, computed offline, software (Figure 4b,c). We decided to use a word-cloud to visualize the titles and keywords of the reviewed 200 papers (Figure 4a) The data table is used to display all information collected and can be filtered by the data selection made by users (Figure 5). All the charts were designed and implemented to link each other through coordinated-view visualization (Section 3.2). Any selection made on one chart would trigger a corresponding filter on other charts based on the current selection. Most importantly, a web map is added to show the geospatial distribution of reviewed/selected papers based on the lead author's institute location. The web map is developed with consideration of cartographic design principles suggested in *idwMapper* in Section 4.2. Major functionalities of the web map include leaflet cluster markers (the points are formed as clusters automatically according to the map scale, as shown in Figure 3a; 50 points are clustered as one point for the USA region when the map is initially loaded if the map is zoomed in, the clustered one point for 50 points will be split into multiple (clustered) points.), informative pop-ups, layer toggling, and inclusion of a mini-map (Figure 3e). The web app includes a short web app demo video: access it on the top-right corner of the web app page. The web app page also has links to a short web app demo video and to the explanations for acronyms and data fields in the data table.

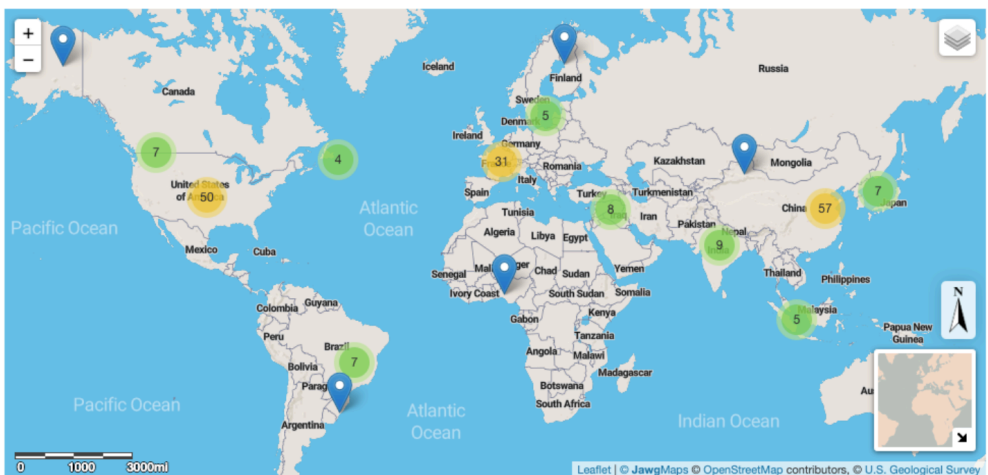

(a)

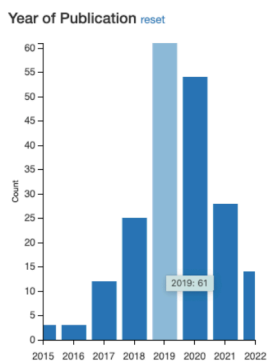 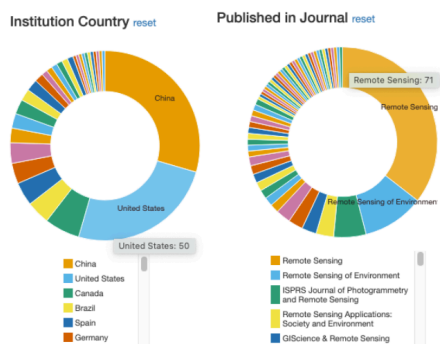 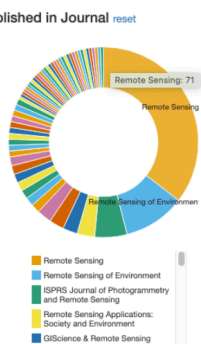 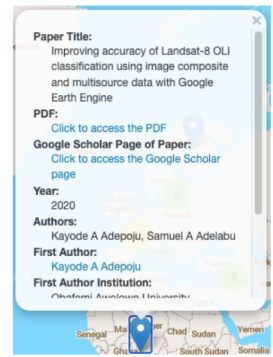

(b)        (c)        (d)        (e)

Figure 3. Geospatial distribution and overview statistics of the 200 reviewed papers. (a) Spatial distribution of reviewed papers based on the first author's institution location, (b) number of published papers by year from 2015 to early 2022, (d) country distribution, and (c) journals the review papers are published in. Note that a freely accessible, interactive version of the map and all charts in this paper can be accessed via our web app tool. A brief web map app demo video is accessible on the web app page (see DATA AND CODES AVAILABILITY STATEMENT).

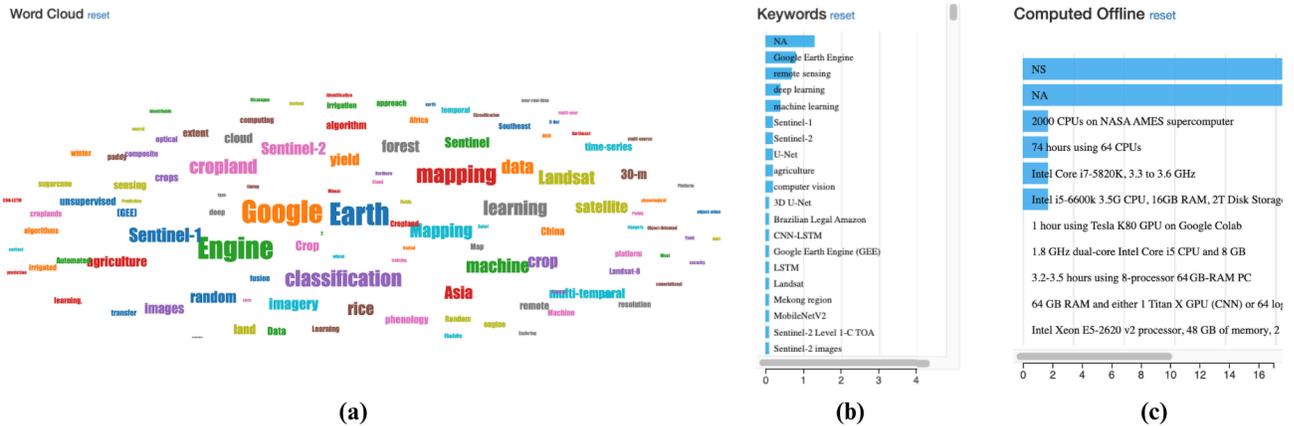

(a)                                (b)                        (c)

Figure 4. Overview of word cloud and row charts (horizontal bar charts). (a) A word-cloud visualization of the titles and keywords of the reviewed 200 papers, (b) row charts with x and y-axis scrolls showing keywords of the reviewed 200 papers, and (c) row charts with x-axis scrolls showing what hardware (e.g., GPU, CPU, RAM, hard disk, runtimes) authors used if they completed part of their analysis offline. Note that a freely accessible, interactive version of the map and all charts in this paper can be accessed via our web app tool.

| Title | Published in Journal | Google Scholar Page of Paper | CitedBy URL | Year of Publication | Authors | Institution Country |
|---|---|---|---|---|---|---|
| 30 m resolution global annual burned area mapping based on Landsat Images and Google Earth Engine | Remote Sensing | Link | 93 | 2019 | Tengfei Long, Zhaoming Zhang, Guojin He, Weili Jiao, Chao Tang, Bingfang Wu, Xiaomei Zhang, Guizhou Wang, Ranyu Yin | China |
| A 30-m landsat-derived cropland extent product of Australia and China using random forest machine learning algorithm on Google Earth Engine cloud computing platform | ISPRS Journal of Photogrammetry and Remote Sensing | Link | 194 | 2018 | Pardhasaradhi Teluguntla, Prasad Thenkabail, Adam Oliphant, Jun Xiong, Murali Krishna Gumma, Russell G Congalton, Kamini Yadav, Alfredo Huete | United States |
| A brave new world for archaeological survey: Automated machine learning-based potsherd detection using high-resolution drone imagery | Journal of Archaeological Science | Link | 43 | 2019 | Hector A Orengo, Arnau Garcia-Molsosa | Spain |
| •••••• | | | | | | |
| A large-scale change monitoring of wetlands using time series Landsat imagery on Google Earth Engine: a case study in Newfoundland | GIScience & Remote Sensing | Link | 27 | 2020 | Masoud Mahdianpari, Hamid Jafarzadeh, Jean Elizabeth Granger, Fariba Mohammadimanesh, Brian Brisco, Bahram Salehi, Saeid Homayouni, Qihao Weng | Canada |
| A Machine Learning-Based Approach for Surface Soil Moisture Estimations with Google Earth Engine | Remote Sensing | Link | 5 | 2021 | Felix Greifeneder, Claudia Notarnicola, Wolfgang Wagner | Italy |

190 selected out of 200 records
Show 10 entries                                                                    Search:

Showing 1 to 10 of 190 entries                            Previous 1 2 3 4 5 … 19 Next

Figure 5. The data table shows the total/selected reviewed papers. The data table has a search box and some columns have clickable links. Note that the table is scrollable horizontally.

## 5.3 iTRELISmap web map app

### 5.3.1 Motivation

TRELIS[18] (Training and Retaining Leaders in STEM - Geospatial Sciences) is a professional development program funded by the National Science Foundation (NSF) for women in the geospatial sciences in higher education. The initiative helps women in geospatial sciences improve in terms of leadership skills, career retention strategies, mentoring training, work-life balance considerations, and practical knowledge. It also encourages them to pass on their expertise to future generations of women in the field. TRELIS has been hosting workshops since 2018, which involve 71 women geospatial scientists whose home institution is located in the United States. TRELIS has been contributing to the broader geospatial communities and, more importantly, emphasizing the women that are overlooked in this GIScience/geospatial data science realm, which has a gender gap. Pavlovskaya (2018) developed the idea that GIS can produce new cartographies and spaces of possibility, leading to geographies of hope and care. We developed iTRELISmap, which primarily serves to highlight female geospatial scientists and educators and promote collaborations among them.

### 5.3.2 Data collection and preprocessing
The data is collected from different academic and social media sites. Academic information is collected from scholars' personal websites, institutional profiles, and their profiles of Google Scholar, ResearchGate, and ORCID. LinkedIn and Twitter profiles are also collected since some scholars use them as their major ways of networking and information exchange. 71 women scholar's degree-related institutions, past worked institutions, and current institutions are collected to connect users based on the scholar's institution. Scholar's research interests and degree focus are also collected to make easier collaboration based on research interests. Detailed data structure and data type explanation are provided in the web app.

### 5.3.3 Web app design and implementation
The implemented iTRELISmap can be accessed at https://geoair-lab.github.io/iTRELISmap/. We have visualized 71 geospatial women scientists whose home institution (when they were awarded TRELIS fellowship) is in the USA. iTRELISmap was designed with the aim of facilitating easy access to information about TRELIS women geographers and encouraging greater connectivity between the audience and women geographers. We carefully considered all relevant information that can be used for our audience to establish connections with these women geographers for academic and career-related purposes. We collected each TRELISer's personal website, email, current institution, location-related information of current institution, research interests, research lab, academic profiles including Google Scholar, ResearchGate, and ORCID profiles, along with social media (LinkedIn and Twitter profiles). Moreover, degree-related institutes, past-worked institutes, and degree focus were also collected to enable networking by institutions and researchers with the same focus. We then selected the appropriate chart types based on the *idwMapper* suggestions detailed in Sections 3.3 and 3.4 and considered the information visualization and cartographic design principle in Sections 4.1 and 4.2. For example, as shown in Figure 6a, the web app shows each scholar's current institution location with markers specially designed for their roles. The map does not tell much more detailed information about each scholar as it is designed to give straightforward spatial distribution information while not adding too much cognitive load to the audience. Yet we added more information in the marker pop-up to allow users to access more detailed information easily from the map if they want (Figure 6e). The photo can be enlarged when hovering over it, and an audio recording of name pronunciation is added. Different profile pages are clickable to give users easy access to information. In the next step, we determined the types of charts to be used: (1) pie charts visualizing cohort number and cohort role, as well as location information (Figure 6b,c,d); (2) bar charts visualizing cohort year and country of the institution (Figure 6); (3) row charts visualizing research interests, degree focus, past worked institutes, and degree-related institutes (Figure 7d); (4) a word-cloud visualization of the research interests (Figure 7e); (5) a data table that displays all information collected and can be filtered by the data selection made by users (Figure 8). Note that all the charts we implemented are interlinked through a coordinated-view visualization approach (Section 3.2). The development of this web map adheres to the cartographic design principles outlined in *idwMapper*, as discussed in Section 4.2. Major functionalities of the web map include leaflet cluster markers, informative pop-ups, layer toggling, and inclusion of a mini-map (Figure 6e). The web app includes short web app demo video: access it on the top-right corner of the web app page. Detailed data structure and data type explanation are provided on this page (https://geoair-lab.github.io/iTRELISmap/dataStructureExplaination.html) of the iTRELISmap web map.

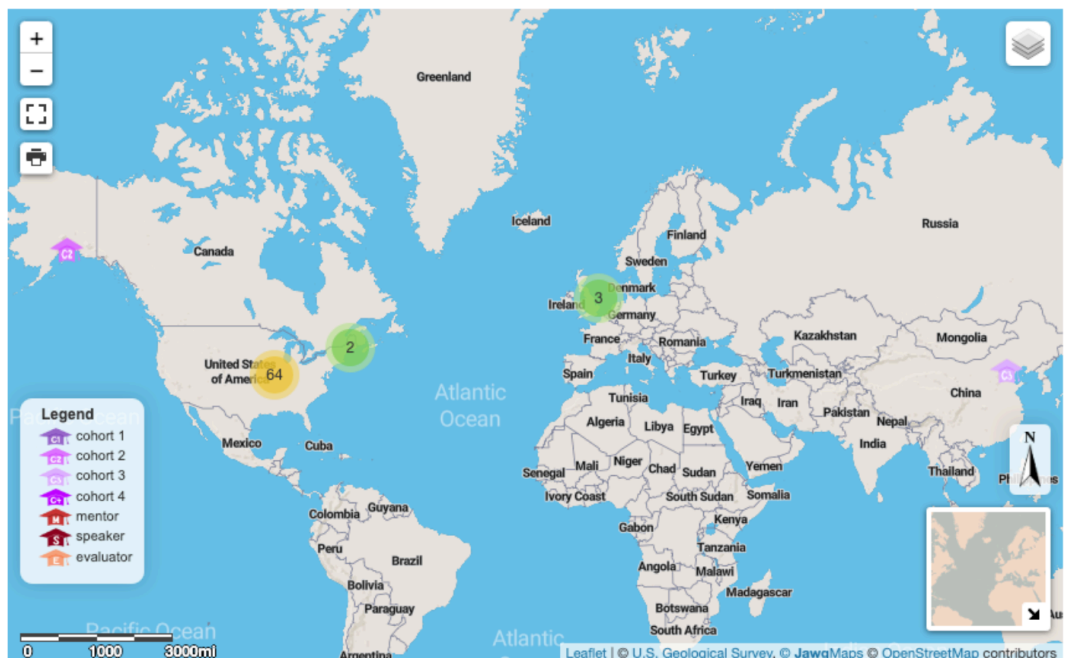

(a)

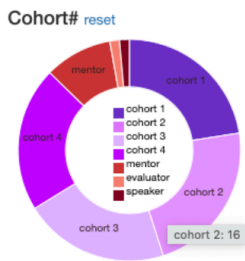

(b)

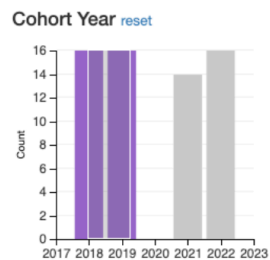

(c)

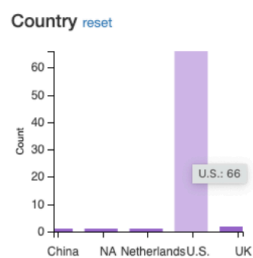

(d)

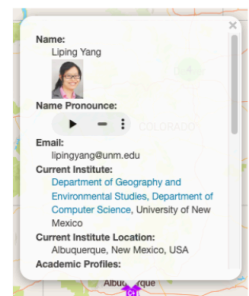

(e)

Figure 6. Geospatial distribution and cohort information of TRELIS. (a) Spatial distribution of TRELIS cohorts, mentors, speakers, and evaluators, based on their current institution locations, (b) TRELISers cohort numbers, where the legend and donut chart piece colors match the legend of the map, (c) cohort year bar chart which has the brushing function enabled. In the figure, 2018

and 2019 are selected to filter other charts so they are in purple and unselected bars are in gray. And (d) the country of TRELISer's current institution. This bar chart does not enable the brushing function, so you can see the slight difference between Figure (c), there is no cross cursor when you hover over the bars, instead there is a click cursor that allows you to select individual bars, (e) a map marker pop-up example of the web app. A brief web map app demo video is accessible on the web app page (see DATA AND CODES AVAILABILITY STATEMENT).

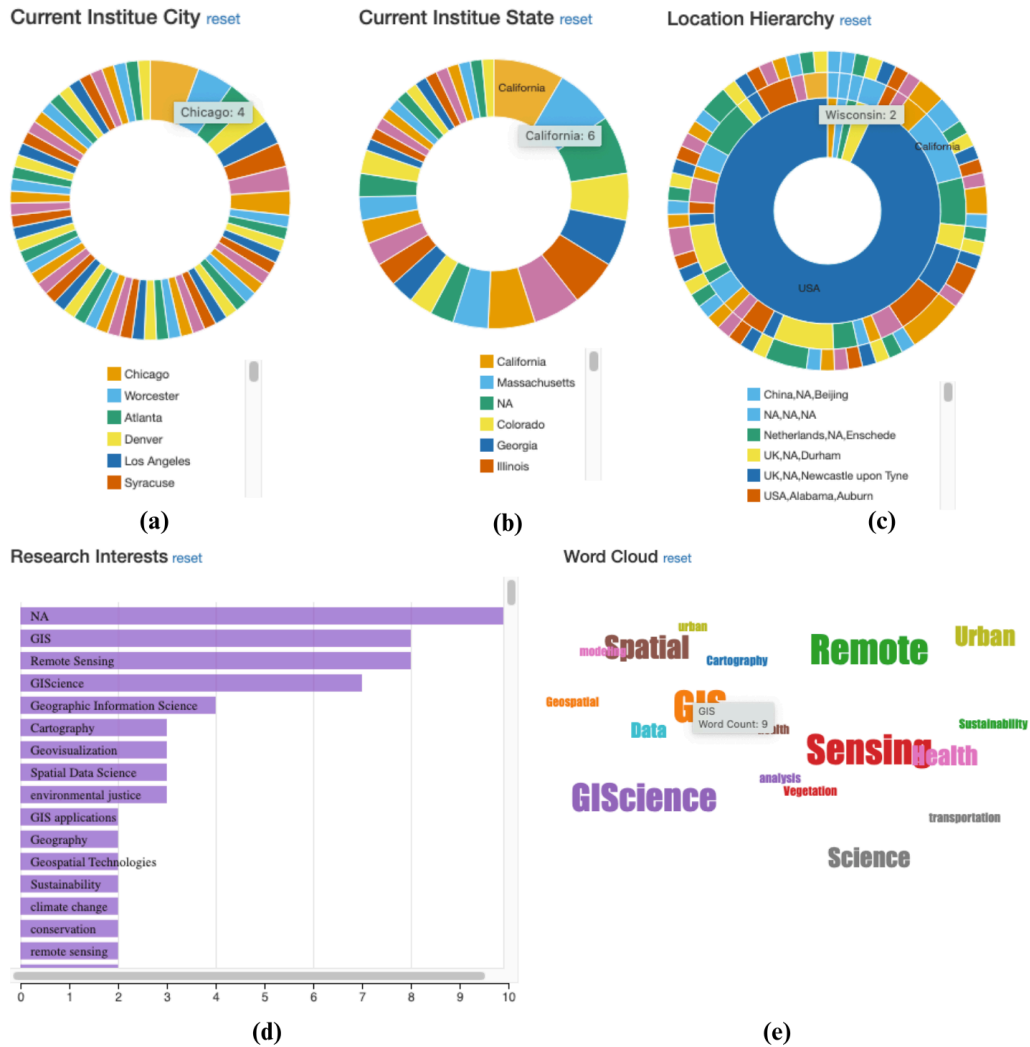

Figure 7. Legend scrollable donut and sunburst charts, row charts, and word cloud. (a) A donut chart visualizes the city of TRELISer's current institution, (b) a donut chart visualizes the state of TRELISer's current institution, and (c) the sunburst chart (hierarchy of location chart) is the merged chart of the city and state charts to help users get the location distribution easily using the hierarchy relationships. (d) A row chart shows TRELISer's research interests, and (e) a word cloud of the research interests of all TRELISers.

71 selected out of 71 records

Show [10 ▼] entries                                                                 Search: [_________]

| Name | TRELIS Role | Cohort# | Cohort Year | TRELIS Workshop Location | Personal Website | Email | Current Institue Name | Current Institue Department | Current Institue City | Current Ins |
|------|-------------|---------|-------------|--------------------------|------------------|-------|------------------------|------------------------------|------------------------|-------------|
| Alicia Cowart | cohort | cohort 4 | 2022 | Blue Mountain Lake, New York | Link | alicia.cowart@ucdenver.edu | University of Colorado, Denver | Geography and Environmental Sciences (GES) | Denver | Colorado |
| Allison Bailey | cohort | cohort 1 | 2018 | Madison, Wisconsin | Link | allison.bailey@ung.edu | University of North Georgia | Institute for Environmental & Spatial Analysis | Atlanta | Georgia |
| Amy Frazier | cohort | cohort 1 | 2018 | Madison, Wisconsin | Link | Amy.Frazier@asu.edu | Arizona State University | School of Geographical Sciences & Urban Planning | Tempe | Arizona |
| • • • • • • | | | | | | | | | | |
| Brandi Gaertner | cohort | cohort 4 | 2022 | Blue Mountain Lake, New York | Link | bah6009@psu.edu | Pennsylvania State University | Department of Geography | University Park | Pennsylvani |

Showing 1 to 10 of 71 entries                                      Previous [1] 2 3 4 5 … 8 Next

Figure 8. The data table shows the total/cross filtered TRELISer(s). The data table has a search box and some columns have clickable links. Note that the table is scrollable horizontally, so our users have access to all the information about TRELISers.

### 5.4 iWURanking web map app

### 5.4.1 Motivation and data collection

University ranking websites have a huge volume of visitors (students and scholars seeking target universities for further education and work opportunities) (Sohail et al., 2021). Academic Ranking of World Universities (ARWU), Quacquarelli Symonds (QS), and Times Higher Education (THE) are the three primary university ranking websites among many others (Sayed, 2019). Our iWURanking applied *idwMapper* aims to visualize the dataset of the QS university ranking. The QS ranking receives approval from the International Ranking Expert Group (IREG). QS designed its world university rankings to assess performance according to what it believes to be critical aspects of a university's mission: teaching, research, nurturing employability, and internationalization. iWURanking aims to help students and scholars around the world gain knowledge and insights about the universities to help them make wise decisions for their future institution of higher education and/or job-seeking through geospatial big data and geovisualization. Note that we use the QS university ranking dataset for a demonstrative purpose, and our iWURanking *idwMapper* case study code implementation can be easily adapted to represent other university ranking data such as Times Higher Education (THE). QS university ranking data and Wikipedia data (including university coordinates; Wikipedia has coordinates information for most of the universities through GeoHack) are collected through web scraping using Python. The data file collected using web scraping provides rankings for approximately 1500 universities from 2019 to 2024, based on six factors that were considered by QS for their ranking methodology. These six factors include academic reputation, employer reputation, faculty to student ratio, number of citations per faculty, international faculty, and international students. Both the rank and score are collected for those six factors. Detailed data structure and data type explanation are provided in this web page (https://geoair-lab.github.io/iWURanking-WebApp/dataStructureExplaination.html) on the iWURanking web map app.

### 5.4.2 Web app design and implementation

Our implemented iWURanking web app can be accessed at
[https://geoair-lab.github.io/iWURanking-WebApp/index.html](https://geoair-lab.github.io/iWURanking-WebApp/index.html). We have visualized 1498
universities recorded on QS Ranking (from 2019 to 2024). There are six factors that determine
the QS world university ranking (see Section 5.4.1). These categories serve as a comprehensive
framework for evaluating various aspects of university performance, ensuring a holistic
understanding of the institution's strengths and areas of focus. By considering these indicators,
individuals can make decisions regarding their educational journey and future prospects. We
collected scores and rankings of these six indicators and the overall scores and rankings of each
university listed on QS Ranking. QS Ranking also had some basic classification information
such as university type (private or public), research output, university size, university age, and
university subject range. Those types of information are also used to create charts. Next, after
analyzing the gathered data, we followed the design principles recommended in *idwMapper* to
select the most appropriate charts for visualizing the data. By selecting the most relevant charts
based on the introduction and description in Sections 3.3 and 3.4, we aimed to enhance the user
experience and facilitate effective information communication to the intended audience.

   We then finalized the chart types, which are detailed below. (1) Pie charts visualize basic
classification information including university type (private or public), research output, and
university size. These charts can be useful for users to do a filter to get general information about
the universities (Figure 9). (2) A web map shows the geospatial distribution of the universities
that is interlinked with other charts, is developed with consideration of information visualization
and cartographic design principles advised in *idwMapper* in Sections 4.1 and 4.2. The web map
enables the Leaflet cluster marker map function, pop-ups with information, switch layers, and
mini-map (Figure 9a). (3) Bar charts visualize the continent where the university is located
(Figure 9b). (4) Row charts visualize the country and city where the university is located. Row
charts are appropriate for country and city charts, especially city charts since there is a large
number of cities and it causes too many tiny invisible pie slices (Figure 9c,d). (5) A sunburst
chart is used to show the hierarchy relation of city and country, and continent that help users
easily access the location  (see Figure 9e). (6) The range/focus chart pair is used to visualize six
indicator scores for the year 2019 to 2024 (Figure 10). The focus chart should give users the
option to brush a certain range, and it displays the corresponding range in the range chart with
more detailed information. Charts are displayed on the page with a collapsible function so that by
clicking the button, the audience can toggle between showing and hiding the collapsible content.
This avoids cognitive overload. (7) A data table that displays all information collected and can be
filtered by the data selection made by users (Figure 11). We also created a brief web map demo
video; the video link is accessible at the web app page (top-right corner); acronyms that are used
in the data table of the web app, as well as explanations for each data field and chart (also in the
top-right corner.

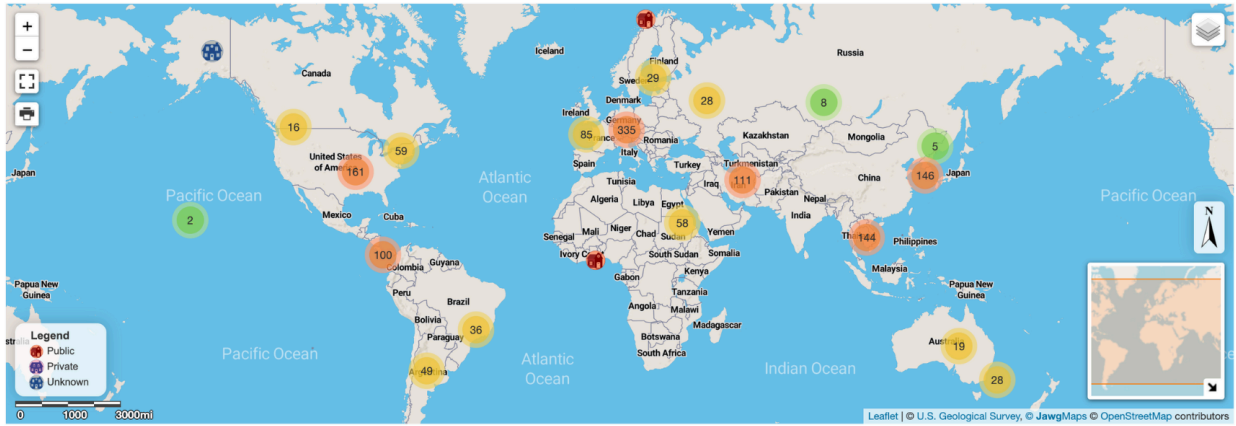

**(a)**

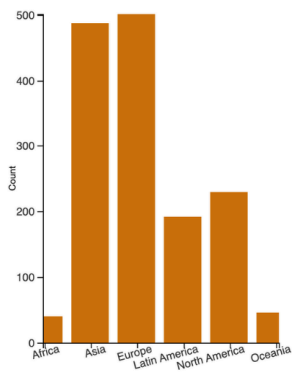

**(b)**

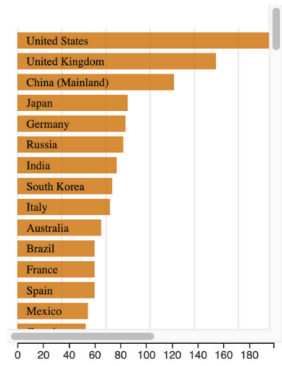

**(c)**

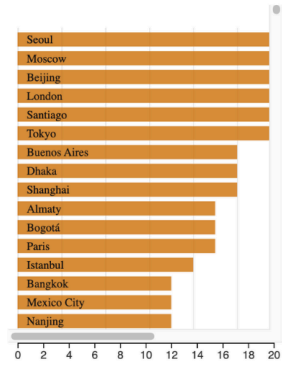

**(d)**

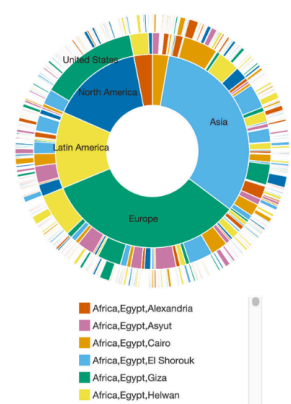

**(e)**

Figure 9. Geospatial distribution and location statistics of the recorded universities. (a) Spatial distribution of universities, (b) university continent bar chart, (c) university country row chart, (d) university city row chart (e) a hierarchy of university location. A brief web map app demo video is accessible on the web app page (see DATA AND CODES AVAILABILITY STATEMENT).

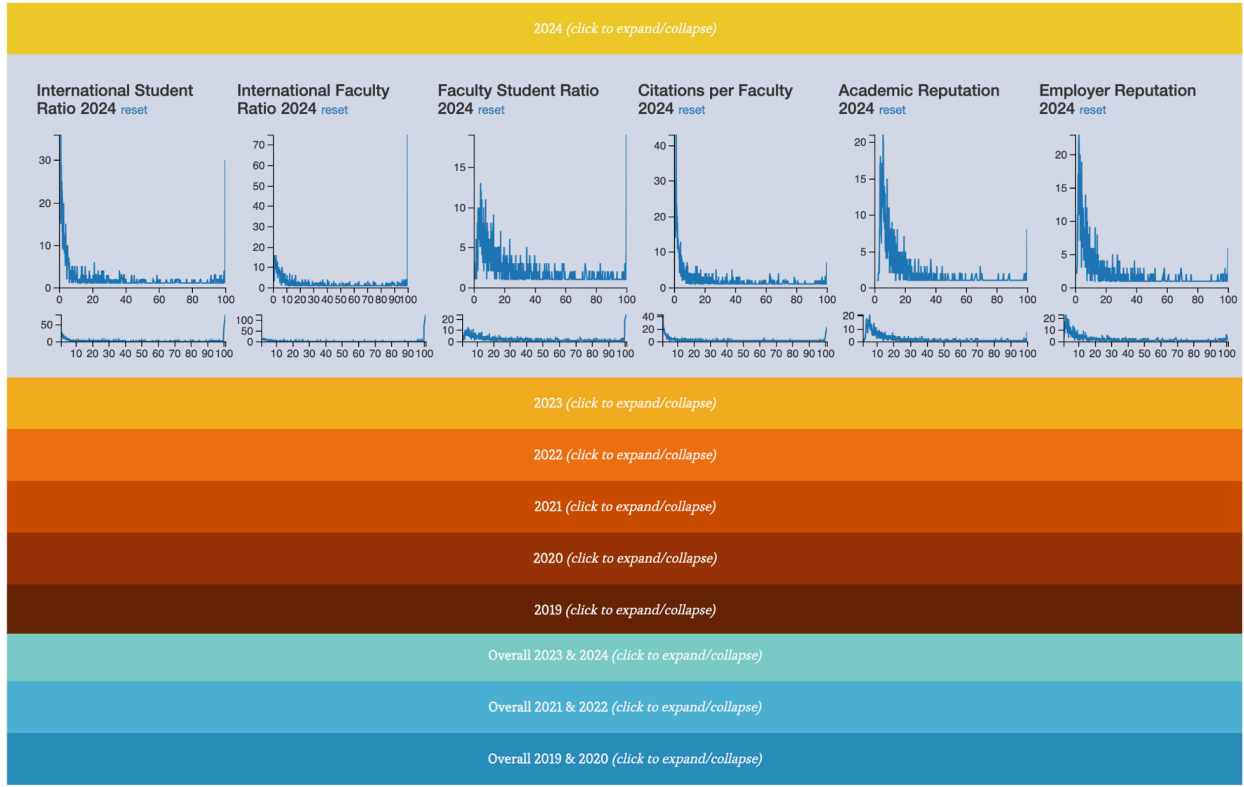

Figure 10. Overview of the score of six indicators of rankings. The zoom and focus line chart pairs visualize the scores of six indicators from 2019 to 2024, respectively, and the overall score from 2019 to 2024 (every two adjacent year are grouped and displayed under one button for space efficient consideration, as the overall ranking has less data compared with the score of six indicators charts mentioned above). Note that each zoom and focus line chart combination has a main chart (focus chart, the one on top) and a range chart (also called zoom chart, the one below the focus chart). Zoom and focus lets the users zoom in and focus on a selected range of data in the charts. For example, if a user wants to focus on universities that have an overall score from 90 to 100, they could move the brush to select the range of 90 to 100 in the zoom chart, the focus chart will then zoom in on these selected scores. Charts for different years are displayed on the page with a *collapsible* function to avoid clutter and to reduce the user's mental information overload., a user can toggle between showing and hiding the collapsible content by clicking the corresponding year button. In this figure, the content of 2024 is expanded and other contents are collapsed and thus being hidden.

Figure 11. The data table shows the total/selected universities. The data table has a search box and some columns have clickable links to extend the usability of the web map.

## 6. CONCLUSION AND DISCUSSION

We proposed and developed an interactive and data-driven web mapping framework named *idwMapper,* which can be used to develop professional web map applications towards sensing high-dimensional geospatial big data in a diverse range of domains. *idwMapper* provides an integration of theoretical and practical guidelines for users from different backgrounds and disciplines to create professional web mapping applications without too much effort in terms of learning web mapping knowledge and cartographic theory. We then designed and implemented three web mapping tools to demonstrate how *idwMapper* can be applied to different domains and problems, with cartographic and information visualization principles implied. Among the three case studies, the iLit4GEE-AI web map successfully helps users quickly filter out published articles that meet their interests. We hope it will become a standard to support researchers in any domain to efficiently write up their literature review articles to advance their fields. *idwMapper* can also be promisingly used to develop EDA tools for ML/statistics projects. It can also help researchers and practitioners easily and intuitively spot data errors. For example, when we developed iLit4GEE-AI, it helped us notice a considerable number of errors when entering the data, such as duplicate entries, errors in data columns, and the non-consistent spelling of the same term. Our *idwMapper* framework works well for point-based data with high dimensions, demonstrated in Section 5 with three different use cases. We expect the framework will work for area-based data with high dimensions, but for line-based data, we will extend our framework in a future version to include representing and analyzing linear features such as roads, bird migration, and airline flows.

## ENDNOTES

1. https://www.esri.com/en-us/arcgis/products/arcgis-dashboards/overview
2. https://shiny.posit.co/
3. https://voila-gallery.org/
4. https://plotly.com/dash/
5. https://streamlit.io/
6. https://panel.holoviz.org/

7. http://bokeh.org/
8. https://colorbrewer2.org/
9. https://pilestone.com/pages/color-blindness-simulator-1
10. https://d3js.org/
11. https://square.github.io/crossfilter/
12. https://dc-js.github.io/dc.js/
13. https://leafletjs.com/
14. https://datatables.net/
15. https://jquery.com/
16. https://getbootstrap.com/
17. https://geoair-lab.github.io/iLit4GEE-AI-WebApp/index.html
18. https://www.ucgis.org/trelis

**DATA AND CODES AVAILABILITY STATEMENT**

The three case study web apps accompanying this paper are all open-source and can be found at the following links:

**iLit4GEE-AI web map:** https://geoair-lab.github.io/iLit4GEE-AI-WebApp/index.html (Demo video link: https://www.youtube.com/watch?v=8iPxjOZdV6k)

**iWURanking:** https://geoair-lab.github.io/iWURanking-WebApp/index.html (Demo video link: https://www.youtube.com/watch?v=0XbgmNGlfWQ)

**iTRELISmap:** https://geoair-lab.github.io/iTRELISmap/index.html (Demo video link: https://www.youtube.com/watch?v=kLYVL2HqWqU)

**ACKNOWLEDGMENTS**

This material is partly based upon work supported by TRELIS Carolyn Merry Mini-Grant (TRELIS is an US NSF funded program, grant no. 1660400), and supported by the funding support from the College of Arts and Sciences, as well as the WeR1 SuRF program (2021 and 2022) at University of New Mexico. The authors are grateful to Junhuan Yang and Gokul Padmanaban for some of their helpful discussions related to the iWURaking web app. The authors are also grateful to the editors and reviewers for their useful suggestions.

**CONFLICT OF INTEREST STATEMENT**

No conflict of interest was reported by the authors.